\documentclass[superscriptaddress,aps,twocolumn]{revtex4}
\usepackage{braket}
\usepackage{graphicx}
\usepackage{chemfig}
\usepackage{amsmath}
\usepackage{amsfonts}
\usepackage{physics}
\usepackage{chngcntr}
\usepackage{float}
\usepackage{comment}
\usepackage[normalem]{ulem}
\usepackage{soul}
\usepackage{hyperref}
\usepackage{comment}
\usepackage{array}

\DeclareMathOperator{\Ai}{Ai}
\usepackage{svg}
\usepackage{float}

\DeclareSymbolFont{matha}{OML}{txmi}{m}{it}
\DeclareMathSymbol{\varv}{\mathord}{matha}{118}

\definecolor{OliveGreen}{rgb}{0,0.6,0}

\begin{document}
\title{Ring-Shaped Linear Waves and Solitons in a Square Lattice of Acoustic Waveguides
}

\author{I.~Ioannou Sougleridis}  
\affiliation{Laboratoire d’Acoustique de l’Universit\'{e} du Mans (LAUM), UMR 6613, Institut d'Acoustique - Graduate School (IA-GS), CNRS,  Le Mans Universit\'{e}, France}
\affiliation{Department of Physics, National and Kapodistrian University of Athens, University Campus, GR-157 84 Athens, Greece}
\author{O.~Richoux}  
\affiliation{Laboratoire d’Acoustique de l’Universit\'{e} du Mans (LAUM), UMR 6613, Institut d'Acoustique - Graduate School (IA-GS), CNRS,  Le Mans Universit\'{e}, France}
\author{V.~Achilleos}  
\affiliation{Laboratoire d’Acoustique de l’Universit\'{e} du Mans (LAUM), UMR 6613, Institut d'Acoustique - Graduate School (IA-GS), CNRS,  Le Mans Universit\'{e}, France}
\author{G.~Theocharis}  
\affiliation{Laboratoire d’Acoustique de l’Universit\'{e} du Mans (LAUM), UMR 6613, Institut d'Acoustique - Graduate School (IA-GS), CNRS,  Le Mans Universit\'{e}, France}
\author{D. J.~Frantzeskakis}  
\affiliation{Department of Physics, National and Kapodistrian University of Athens, University Campus, GR-157 84 Athens, Greece}

\begin{abstract}
We study the propagation of both low- and high-amplitude ring-shaped sound waves in a 2D square lattice of acoustic waveguides with Helmholtz resonators. We show that the inclusion of the Helmholtz resonators suppresses the inherent anisotropy of the system in the low frequency regime allowing for radially symmetric solutions. By employing the electroacoustic analogue approach and asymptotic methods we derive an effective cylindrical Korteweg de Vries (cKdV) equation.  Low-amplitude waveforms are self-similar structures of the Airy function profile, while high-amplitude ones are of the form of cylindrical solitons. Our analytical predictions are corroborated by results of direct numerical simulations, with a very good agreement between the two. 
%
%
\end{abstract}
\maketitle

\section{Introduction}

Over the past years, structured materials have been widely developed to manipulate wave motion. In that respect, acoustic metamaterials, namely structured materials made of resonant building blocks, play an important role in the design of various classical wave systems. Earlier studies on acoustic metamaterials based on acoustic waveguides incorporating resonant elements (e.g., Helmholtz resonators \cite{sugimoto_dispersion,sugimoto_dispersion_2} or  
quarter-wavelength resonators \cite{bradley1,bradley2,bradley3}) paved the way for 
a variety of important applications. These include  acoustic diodes \cite{diode}, perfect absorbers \cite{vassos_nonlinear,perfect_absorption_1,metasurfaces_1}, acoustic lenses for sub-diffraction imaging \cite{lens}, acoustic sound focusing based on gradient index lenses \cite{gradient,gradient_2}, acoustic topological systems \cite{Ma,coutant2021acoustic} acoustic cloaking \cite{cloaking_1,cloaking_2,cloaking_3,cloaking_4}, bifurcation-based acoustic switching and
rectification \cite{bifurcation}, and so on. 


The above studies refer to one-dimensional (1D) settings and deal with linear wave phenomena. Nevertheless, there exist a number of works where nonlinear wave phenomena in 1D acoustic metamaterials were investigated. 
Relevant studies were motivated by the seminal works of Sugimoto and coworkers on acoustic solitons in air-filled waveguides with Helmholtz resonators \cite{sugimoto_prl,sugimoto_review} (see also the relevant works \cite{wave_motion,vassos_soliton}). 
%
%
As shown in these works, acoustic solitons obey an effective Korteweg-de Vries (KdV) equation \cite{ablowitz} in the long-wavelength and small-amplitude limits.  
%
%
Other settings, involving waveguides loaded with arrays of elastic membranes or side holes, were also predicted to support acoustic envelope solitons, governed by an effective nonlinear Schr\"{o}dinger (NLS) equation ~\cite{kinezoula_dark,kinezoula_gap}. 
%
%

In addition to the aforementioned studies in 1D settings, there exist also many works on two-dimensional (2D) and 3D-dimensional (3D) acoustic metamaterials. These works have shed light on wave phenomena relevant to topological insulators \cite{2dacoustic,mobius2d}, acoustic analogues of optical fibers \cite{analogue_fiber}, acoustic hyperbolic materials \cite{hyberbolic_acoustic}, and others. Nevertheless, it should be pointed out that the above mentioned studies refer to linear acoustic waves, while 
works on nonlinear wave phenomena in airborne 2D or 3D acoustic metamaterials are rather limited. In fact, such works are basically devoted to elastic and mechanical metamaterials \cite{elastic_metamaterials,deng2021nonlinear}, with one of the main goals being the investigation of the formation, propagation and collision of solitary pulses 
\cite{tournat_soliton,vector_mechanical}.


In this work, our scope is to investigate nonlinear wave phenomena in a 2D acoustic network, composed of a square lattice of waveguides,  
loaded with Helmholtz resonators at the junctions' locations (see also Refs.~\cite{depollier,acousticgraphene} for studies in similar setups). We find that, in the long-wavelength limit, wave dynamics in the acoustic waveguide network (which may be viewed as a 2D generalization of the setting studied in Refs.~\cite{sugimoto_prl,sugimoto_review,wave_motion,vassos_soliton}) 
can be described by an effective cylindrical KdV (cKdV) equation (which is the 2D, radially-symmetric counterpart of the usual 1D KdV equation). The cKdV description that we establish here, allows us to predict both low-amplitude (linear) ring-shaped waves, as well as high-amplitude (nonlinear) waveforms. 
%
%
Much like the 1D KdV, the cKdV equation is a universal model describing the evolution of ring-shaped nonlinear waves and solitons 
in various physical contexts, including water waves \cite{johnson1999note,mannan2015ring,radial_boussinesq}, plasmas \cite{KO,plasma_cylindrical}, square electrical lattices \cite{stepanyants1981}. 

A brief presentation of our methodology and findings, along with the description of the organization of the paper, are as follows. In Sec.~II, we present our setting, namely the square acoustic network, with and without the Helmholtz resonators. We  show that, in the long-wavelength regime, the resonators suppress the inherent anisotropic dispersion of the network. In addition (also in Sec.~II), we introduce the electroacoustic analogy (EA) through the fluid conservation laws. This approach leads to an effective 2D Boussinesq equation which, in the limit of large-radii and small-amplitude (weakly nonlinear) waveforms, is then reduced to the cKdV equation. Next, in Sec.~III, we present linear and nonlinear ring-shaped solutions of the effective cKdV. The former are self-similar solutions of the linearized cKdV exhibiting an Airy-function profile, while the latter are cylindrical solitons. In the same Section (Sec.~III) we present results of direct numerical simulations, for both linear and nonlinear ring-shaped waves in the acoustic network. 
Finally, in Sec.~IV, we summarize our findings and propose feature research directions.


\section{The Model and its analytical consideration}

\subsection{Acoustic network and dispersion relation}

\begin{figure}[tbp!]
\begin{center}
    \includegraphics[width=0.45\textwidth]{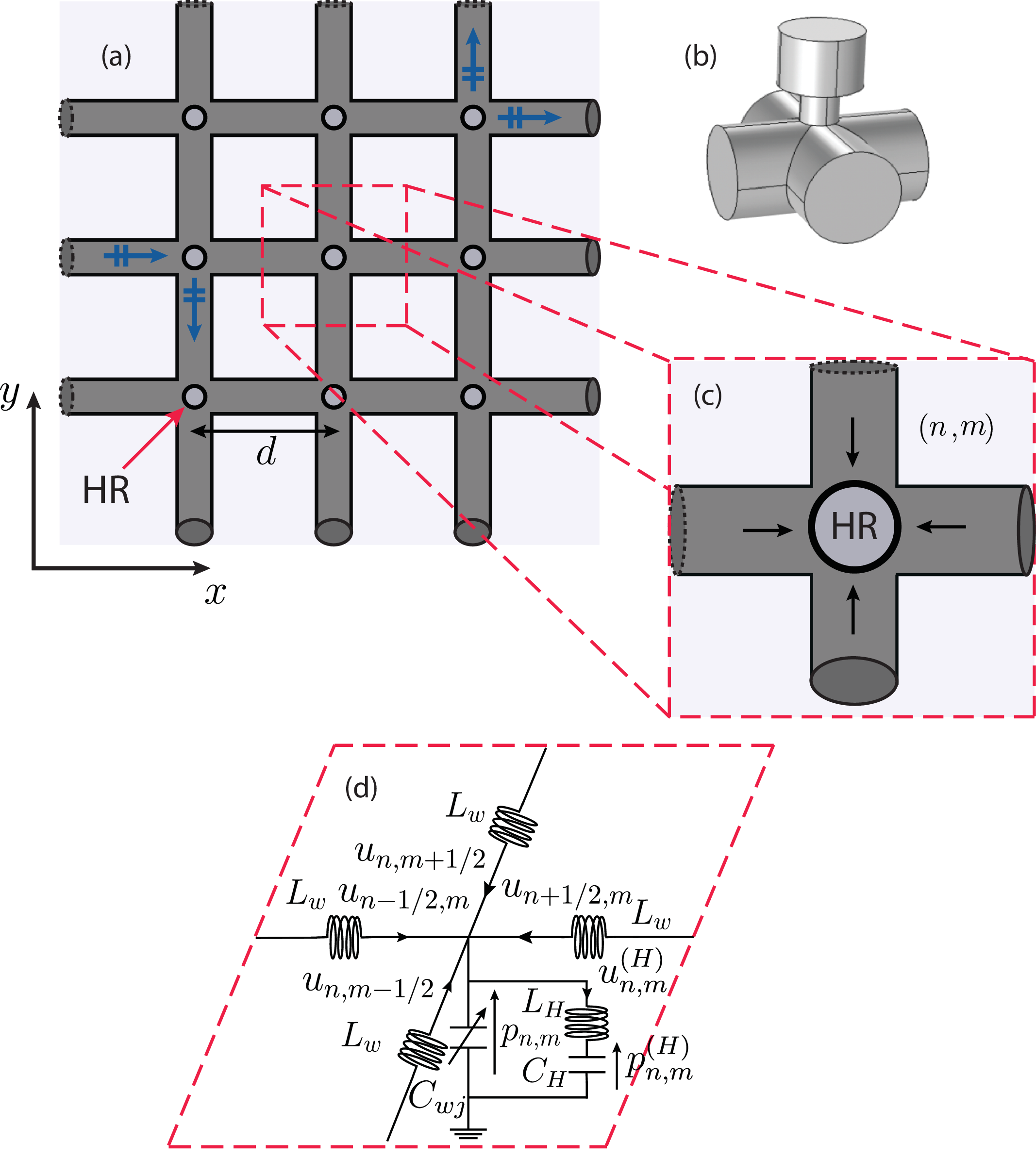}
  \caption{Square acoustic network with Helmholtz resonators. (a) Physical setup. (b) A 3D view of the unit cell. (c) A contour of the unit cell. (d) Unit cell of the electroacoustic analogue transmission line. 
  \label{resonators system}}
  \end{center}
\end{figure}

We consider an acoustic network, composed of simply connected waveguides, of cross-section $S_w$, arranged in a square lattice with lattice distance $d$ ---see Fig.~\ref{resonators system}. We assume that the waveguides are filled with air, which will be  treated as an ideal fluid; thus, viscosity and other dissipative
terms will not be taken into account. 

\begin{figure}[tbp!]
\begin{center}
\includegraphics[width=0.475\textwidth]{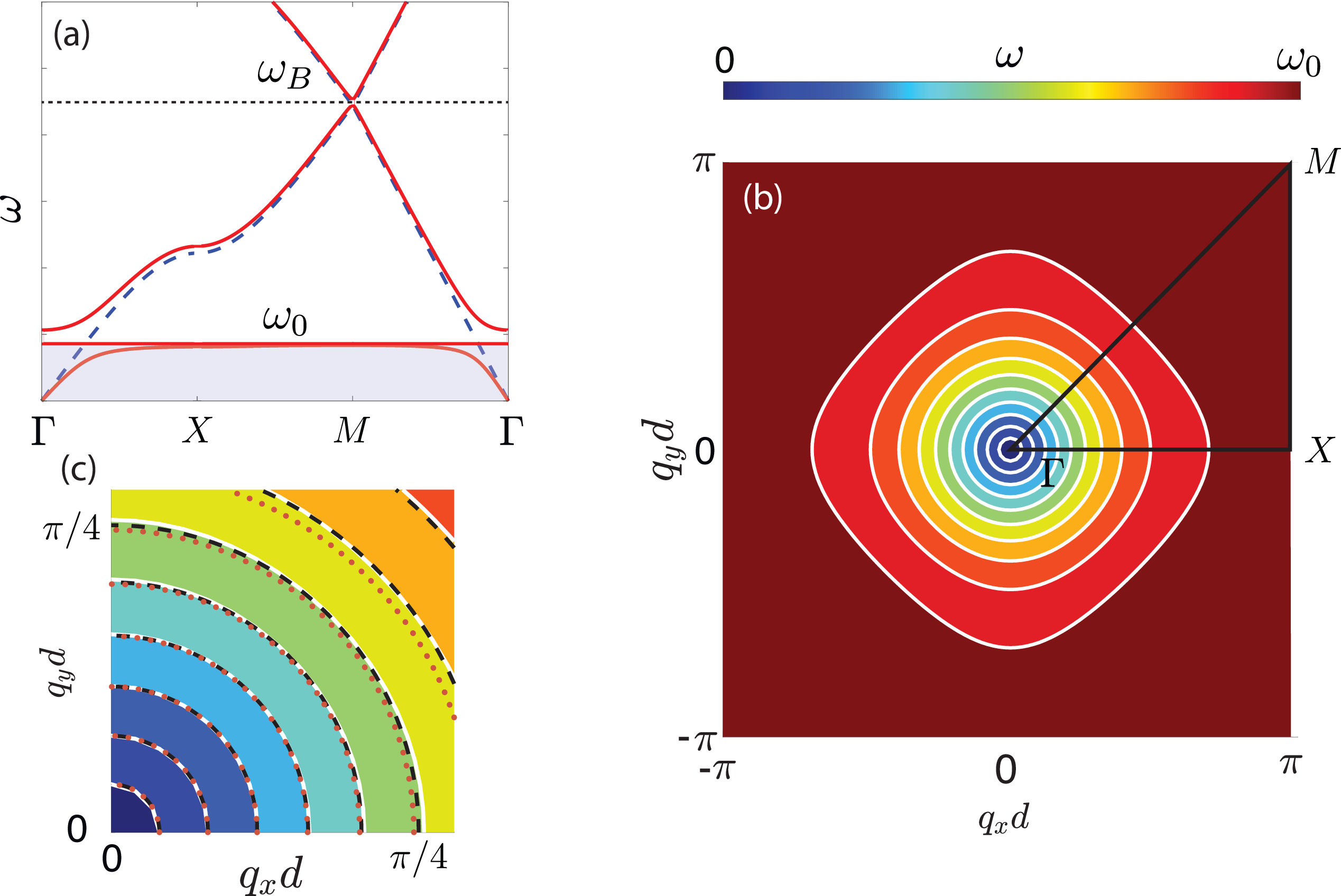}
\caption{Band structure of the system. (a) Dispersion relation of the acoustic network with [solid (red) lines] and without [dashed (blue) lines] Helmholtz resonators, along the axis of high symmetry. The dotted (black) line corresponds to the Bragg frequency $\omega_B$, while the (light grey) rectangle depicts the longwavelength regime, $\omega<\omega_0$ (for the network with resonators). (b) Contour of the first branch  ($\omega<\omega_0$) of the dispersion relation; solid  (white) lines depict isofrequency contours, while the triangle $\Gamma X M$ denotes the first irreducible Brillouin zone. (c) Zoom of the dispersion contour, where the isofrequency lines of the TMM Eq.\,\eqref{TMM resonators} are compared with the Boussinesq model [dashed (black) lines] and the cKdV longwavelength limit [(red) circles]. \label{dispersion}} 
  \end{center}
\end{figure}

In the long-wavelength approximation, i.e., for wavelengths much larger than the waveguide's cross section, we assume that only the plane mode is propagating in each waveguide; this will be referred to as the ``monomodal approximation'' hereafter. Hence, in the linear regime, one can model wave propagation between the junctions of the network with the one-dimensional (1D) Helmholtz equation. As shown in Refs.~\cite{depollier,topology_2D,acousticgraphene}, the transfer matrix method (TMM) may be used to derive a set of equations for the pressure in the square network. These equations lead to the dispersion relation (see Appendix~\ref{appendixtmm} for details):
\begin{align}
\cos \left(q_x d\right)+\cos \left(q_y d\right)
=2 \cos (k d), 
\label{TMM no resonators}
\end{align}
where $k=\omega/c_0$ (where $\omega$ is the frequency and $c_0$ is the linear speed of sound), while $q_x$ and $q_y$ are the Bloch wavenumber components along $x$ and $y$ respectively. This dispersion relation is depicted in Fig.~\ref{dispersion}(a) ---see (blue) dotted line. 
%
Importantly, the dispersion relation exhibits different dispersive behavior depending on the direction of propagation. In particular, as shown in Fig.~\ref{dispersion}(a), along the diagonal $\Gamma M$ 
of the first irreducible Brillouin zone the network exhibits \textit{dispersionless} propagation, while along the horizontal $\Gamma X$ direction the system features a \textit{strong} dispersion. 

In this work, we are interested in radially symmetric waveforms, namely cylindrical pulses that may propagate within the network, which do not feature an angle dependent amplitude and width. Thus, we need to adjust the long-wavelength dispersive characteristics of the square lattice, namely suppress the anisotropy, to achieve isotropic propagation. This can be done by side-loading Helmholtz resonators (HRs) to each node of the square network, which introduce local resonances to the network.
%
%
%

To model the dispersion of the network with the HRs, we will again employ the TMM, and derive a set of discrete equations for the pressure field. This way, we arrive at the following dispersion relation (see Appendix \ref{appendixtmm}),
\begin{equation}
\cos \left(q_xd\right)+\cos \left(q_y d\right)=2\cos \left(k d\right)+\mathrm{i}\frac{Z_w}{2Z_{HR}}\sin \left(k d\right), \label{TMM resonators}
\end{equation}
which is identical to Eq.~\eqref{TMM no resonators}, but also incorporates the last term in the right-hand side;  
here, $Z_w$ and $Z_{HR}$ denote the characteristic impedance of the waveguide segments and the entrance impedance of the HRs, respectively (see Appendix \ref{appendixtmm}). 

The corresponding band structure of the acoustic network is illustrated in Fig.~\ref{dispersion}. In particular, Fig.~\ref{dispersion}(a) depicts the dispersion relation Eq.\,\eqref{TMM resonators} of the acoustic network without [dashed (blue) line] and with [solid (red) line] the HRs, along the axis of high symmetry of the irreducible Brillouin zone. The Bragg frequency $\omega_B=\pi c_0/d$ of the network is illustrated by the dotted (black) line, while the transparent rectangle depicts the low frequency regime, i.e., $\omega<\omega_0$, where $\omega_0$ is the resonance frequency of the HR. Figure~\ref{dispersion}(b) shows the contour plot ---with solid (white) lines illustrating isofrequency contours--- of the first branch ($\omega<\omega_0$) [(light grey) rectangle] of the dispersion relation Eq.~\eqref{TMM resonators}; furthermore, the triangle $\Gamma X M$ denotes the first irreducible Brillouin zone. A zoom of the isofrequency contour in the longwavelength regime, is shown in Fig.~\ref{dispersion}(c). Here, the isofrequency lines of the TMM Eq.
\eqref{TMM resonators} are also compared with the ones corresponding to the effective models that we will introduce below, namely the Boussinesq model [dashed (black) line] and its cKdV low frequency limit [(red) circles]. It can readily be seen that the inclusion of the resonators induces dispersion in the low frequency regime, in the region where the square network is \textit{dispersionless}, through a hybridization gap originated by the HR resonance $\omega_0$. 
Ultimately, the resonators not only introduce dispersion to the system, but in the first branch of the dispersion relation,  $\omega<\omega_0$, they also effectively suppress the inherent anisotropic dispersive behavior of the square network.

Note that throughout this study, we use the geometrical parameters based on experimental realizations of a 1D waveguide sideloaded with a periodic array of HRs in \cite{wave_motion,vassos_soliton}. In particular, for the waveguide segments, we use $d=10$\,cm as the lattice distance and  $r_w=2.5$\,cm for the radius.  As for geometrical characteristics of the HR, namely, the radius of the neck and the cavity, the length of the neck and the height of the cavity we use
$r_n=1$\,cm, $r_c=2.15$\,cm, $l=2$\,cm and $h=16.5$\,cm respectively. These values fix the resonance frequency of the HR at $\omega_0=2\pi\times441$\,rad/s in physical units.

\subsection{Electroacoustic analogue}

The monomodal approximation that was adopted in the previous section to characterize the dispersive properties of the network allows us to employ a quasi-1D description of our setting. In particular, we consider that each waveguide segment (in between successive junctions) is 1D, so that the governing fluid equations assume a quasi-1D form. This way, in each  waveguide segment, along the $x$- or $y$-direction, the mass conservation (continuity equation) and momentum conservation (Euler equation) take the form, respectively,
\begin{eqnarray}
&&\frac{\partial \rho}{\partial t}+\frac{\partial}{\partial \nu}(\rho v_\nu)=0, 
\label{masscons} \\ 
&&\rho \left(\frac{\partial v_\nu}{\partial t}+v_{\nu}\frac{\partial v_\nu}{\partial \nu}\right) =-\frac{\partial p}{\partial \nu},
\label{momcons}
\end{eqnarray}
where, $\nu=x,y$. Here, $\rho=\rho(x,y,t)$ is the density, and $p=p(x,y,t)$ is the pressure (both referring to the whole network), which are connected via the equation of state $p=p(\rho,s)$; here $s$ is the entropy, which hereafter is assumed to be constant. Furthermore, $v_x=v_x(x,t)$ and $v_y=v_y(y,t)$ are the velocity components for a waveguide segment along the $x$- or $y$-direction [see Fig.~\ref{resonators system}(a)]. 

We now consider the unit cell of the system (see Fig.~\ref{resonators system}(c)], which includes a junction connecting four waveguide segments. Under the aforementioned assumptions, each junction may be considered as a single point, for which the conservation of mass takes into account the coupling between the four waveguide segments and the presence of the HR (see, e.g., Ref.~\cite{sugimoto_dispersion,merciernew}), namely:
\begin{align}
& 2\frac{\partial \rho}{\partial t}+\frac{\partial}{\partial x}(\rho v_x)+\frac{\partial}{\partial y}(\rho v_y) = \frac{1}{S_n}\oint \rho v^{(H)}dl, \label{mass fluid}
\end{align}
where $v^{(H)}$ is the velocity component in the neck of the HR, and $S_n$ is the surface area of the neck. The term on the right hand side of Eq.\,\eqref{mass fluid} corresponds to the mass flux through the orifice of the neck of the resonator.

Considering solutions on top of the equilibrium state defined by the density of air $\rho_0$ and atmospheric pressure $p_{\text{atm}}$, we will make use of the substitutions 
$\rho\to\rho_0+\rho$ and $p\to p_{\text{atm}}+p$. 
In addition, assuming that 
$S_n/S_w \ll 1$ and $S_n/S_c \ll 1$  (where $S_w$ and $S_c$ are the cross-sections of the waveguide segments and the HR's cavity, respectively), the right-hand side of Eq.~(\ref{mass fluid}) can be approximated as~\cite{rienstra_nonlinear,sugimoto_dispersion,merciernew} %
$(1/2S_n)\oint \rho v^{(H)}dl ~\approx~ (r_n/S_n) \rho_0 v^{(H)}$, where nonlinear effects inside the HR cavity are neglected. The latter approximations are justifiable, upon substituting the geometrical parameters defined at the end of the previous section.

Next, we employ the quadratic approximation of the equation of state \cite{hamilton,enflo,rudenko}, according to which the density is expressed as the leading-order terms of the Taylor expansion of the pressure, namely:
\begin{equation}
 \rho\approx\frac{p}{c_0^2}-\frac{\gamma-1}{2\rho_0c_0^4} p^2, \label{quadratic}
\end{equation}
where $\gamma$ is the specific heat ratio. We then substitute 
Eq.~(\ref{quadratic}) 
into Eqs.~\eqref{masscons}-\eqref{momcons} and keeping only quadratic nonlinear terms in pressure  \cite{hamilton,enflo}, we  obtain:
\begin{align}
&\frac{1}{c_0^2}\frac{\partial p}{\partial t}
-\frac{\beta_0}{\rho_0c_0^4}\frac{\partial \left(p^2\right)}{\partial t}
+\rho_0 \frac{\partial  v_\nu}{\partial \nu}=0, \label{mass simplified waveguide}  \\
&\rho_0\frac{\partial v_{\nu}}{\partial  t} =\frac{\partial p}{\partial \nu}, \label{momentum simplified}   
\end{align}
where $\beta_0=\gamma+1=1.2$ for air, and we have considered plane progressive waves, with $p=\rho_0 c_0 v_\nu$ (see details in chapter~3 of \cite{hamilton}). 
In addition, approximating again the density in Eqs.~(\ref{mass fluid}) as per Eq.~(\ref{quadratic}), and keeping only quadratic nonlinear terms, we obtain the following mass conservation for the junction:
\begin{eqnarray}
\frac{2}{c_0^2}\frac{\partial p}{\partial t}
-\frac{2\beta_0}{\rho_0c_0^4}\frac{\partial \left(p^2\right)}{\partial t}+
\rho_0 \left(\frac{\partial  v_x}{\partial x}
+\frac{\partial v_y}{\partial y}\right) = \frac{2r_n}{S_n}\rho_0v^{(H)}.\nonumber\\
\label{mass simplified} 
\end{eqnarray}
%

%

To bypass the difficulties in analyzing the system of Eqs.~\eqref{mass simplified waveguide}-\eqref{mass simplified}, we now employ the electroacoustic analogue (EA) proposed for acoustic settings \cite{lissek,acoustic_meta_tl,vassos_soliton,kinezoula_membrane,sougleridis2023acoustic}. In particular, we consider a 2D electrical transmission line, as a long wavelength approximation of the 2D network; the unit cell of the transmission line is depicted in Fig.~\ref{resonators system}(d). The dynamics of this effective transmission line can be described by Kirchhoff current law (KCL) and Kirchhoff voltage law (KVL), that mimic the quasi-1D conservation equations \eqref{mass simplified waveguide}-\eqref{mass simplified}. In particular, according to the EA approach, acoustic flux (pressure) corresponds to current (voltage), with KCL and KVL leading to a discretized version of Eqs.~\eqref{mass simplified waveguide}-\eqref{mass simplified}; the latter will be particularly relevant for our analytical and numerical investigations, as we will see below.   

First, KCL (corresponding to mass conservation) for each junction $(n,m)$ yields
\begin{align}
&u_{n, m-1/2}+u_{n-1/2, m}+u_{n+1/2, m}+u_{n, m+1/2}\nonumber\\
&-u^{(H)}_{n,m}
=\frac{d}{d t}\left[C_{w,j}(p_{n,m})p_{n, m}\right], \label{current waveguide}
\end{align}
where $u_{n\pm 1/2,m\pm 1/2}$ is the acoustic flux per the lattice distance $d$, $u^{(H)}_{{n,m}}$ is the flux per the diameter of the HR's neck $2r_n$, and $p_{n,m}$ is the pressure at each junction. Notice that the discrete pressure field is defined only at the junctions $(n,m)$, while due to the central finite differences, the acoustic flux is defined in the middle of each waveguide in between two consecutive junctions; the latter locations are denoted as $(n\pm1/2,m\pm1/2)$. Finally, $C_{w,j}$ is a pressure-dependent capacitance, which is nonlinear due to the presence of the quadratic term in the equation of state Eq.~\eqref{quadratic}. In particular, $C_{w,j}$ can be  approximated as,
\begin{equation}
    C_{w,j}\approx C_{w,j0}(1-bp_{n,m}),\quad  \nonumber
\end{equation}
where $C_{w,j0}=2dS_w /\rho_0 c_0^2$ is the linear part of the capacitance and $b=2\beta_0/\rho_0 c^2_{0}$ is the nonlinearity coefficient. 
Second, KVL (corresponding to the momentum conservation) for each junction reads:
\begin{align}
&L_w \frac{d}{d t} u_{n, m\pm1/2}=p_{n,m\pm1}-p_{n,m},\label{voltage waveguide1}\\
&L_w \frac{d}{d t} u_{n\pm1/2,m}=p_{n\pm1,m}-p_{n,m}, \label{voltage waveguide2}
\end{align}
with $L_{w}=\rho_0 d/S_w$ being the inductance. 
Furthermore, KVL for the branch corresponding to the resonator located at the $(n,m)$ connection, leads to:
\begin{equation}
    L_{H}\frac{d}{dt}u^{(H)}_{n,m}=p_{n,m}-p^{(H)}_{n,m}, \label{voltage resonator}
\end{equation}
where $L_H=\rho_0l/S_n$ is the resonator's effective inductance, and $p^{(H)}_{{n,m}}$ is the pressure inside the resonator cavity at the $(n,m)$ junction. 
Additionally, the KCL for the resonator branch yields,
\begin{equation}
    C_{H}\frac{d}{dt}p^{(H)}_{{n,m}}=u^{(H)}_{{n,m}} \label{current resonator}.
\end{equation}
As per the assumptions of the previous section (see also Ref.~\cite{vassos_soliton}), the capacitance of the resonator is assumed to be linear, $C_H=V_H\rho_0/ c_0^2$.
Combining the Kirchhoff laws for the HR,  Eqs.\,(\ref{voltage resonator}- \ref{current resonator}), one obtains the following expression for the flux in the resonator's neck:
\begin{equation}
u^{(H)}_{n,m}=\frac{1}{L_{H}}\hat{P}^{-1} \frac{d}{dt}p_{n,m}, \label{operator}
\end{equation}
where  $\hat{P}=\left(d^2/dt^2+1/(L_{H}C_{H})\right)$, with $1/(L_{H}C_{H})=\omega_0^2$ (recall that $\omega_0$ is the HR's resonance frequency).
 By substituting the flux of the resonator,  Eq.\,\eqref{operator}, into the Kirchhoff laws for the junctions, Eqs.\,(\ref{current waveguide}-\ref{voltage waveguide2}), we obtain the following differential-difference equation (DDE) for the pressure $p_{n,m}$,
\begin{align}
   &L_w C_{H } \frac{d^2 p_{n, m}}{d t^2}-\left(1+\frac{1}{\omega_0^2}\frac{d^2}{d t^2}\right) \hat{\delta}_{n, m}^2 p_{n, m} \nonumber \\ 
   &+L_wC_{w,{j0}}\frac{d^2}{d t^2}\left(1+\frac{1}{\omega_0^2}\frac{d^2}{d t^2}\right)\left(p_{n, m}-bp^2_{n, m}\right)=0,\label{nonlinear forth order}
\end{align}
where 
\begin{align}
\delta^2_{n,m}p_{n,m}:=&p_{n+1,m}-2p_{n,m}+p_{n-1,m}\nonumber\\
&+p_{n,m+1}-2p_{n,m}+p_{n,m-1},\nonumber
\end{align}
is the discrete Laplacian. Measuring fluxes, pressures and time in units of $1/\sqrt{L_{w0}}$, $1/\sqrt{C_{w0}}$ and  
$\sqrt{L_{w0}C_{w0}}$ respectively,  
%
where $L_{w0}=\rho_0 d/2S_w$ and $C_{w0}=dS_w /2\rho_0 c_0^2$ are the transmission line elements for a waveguide segment of length $d/2$ (see also discussion in Appendix~\ref{appendix num}), 
Eq.~\eqref{nonlinear forth order} is rewritten in the following dimensionless form:
\begin{eqnarray}
&&\frac{d^2}{dt^2}p_{n,m}-\frac{c^2}{d^2}\left(1-\frac{1}{\omega_0^2}\frac{d^2}{dt^2}\right)\delta_{n, m}^2 p_{n,m}
\nonumber \\
&&+\frac{1}{(\kappa+1)\omega_0^2}\frac{d^4}{dt^4}p_{n,m}
-\frac{b}{\kappa+1}\Bigg(\frac{d^2}{dt^2}\left(p_{n,m}^2\right)
\nonumber \\
&&+\frac{1}{\omega_0^2}\frac{d^4}{dt^4}\left(p_{n,m}^2\right)\Bigg)=0, \label{normalized nonlinear forth}
\end{eqnarray}
where 
\begin{eqnarray}
c^2=\frac{d^2L_{w0}C_{w0}}{L_w(C_{wj,0}+C_H)}, 
\quad  
\kappa = \frac{C_H}{C_{wj,0}}, 
\end{eqnarray}
while we have used the substitutions 
$(L_H C_H)/(L_w C_{w0}) ~\to~ 1/\omega^2_0$ and  
$b\to b/\sqrt{C_{w0}}$. 
%
%
Notice that the pressure $p_{n,m}$, the distance across the lattice and time in the above equation are respectively measured in units of $1.2\times 10^{-5}~{\rm Pa}$, $d=10~{\rm cm}$, and $2.9\times 10^{-5}$ seconds.

\subsection{Continuum approximation -- cKdV equation}

To study analytically the DDE\, (\ref{normalized nonlinear forth}), we resort to the continuum approximation. This will allow us to derive an effective partial differential equation (PDE) for the pressure in the network, which will turn to be the cKdV equation. As we will see, the latter admits radially symmetric solutions that can be supported by the acoustic waveguide network. 

In the long wavelength regime, with $\omega<\omega_0$ (where the field is varying slowly from junction to junction),  
the pressure $p_{n,m}(t)$ can be approximated by a continuum variable, i.e., $p_{n,m}(t) \approx p(x,y,t)$, where   $x=nd$, $y=md$. Consequently,  
the discrete Laplacian can be approximated by a Taylor expansion of $p_{n,m}(t)$ at the junction $(n,m)$ as
\begin{align}
&\delta^2_{n,m}p_{n,m}\approx d^2 \Delta p, \quad \mbox{ with } \Delta \equiv \partial_x^2 +\partial_y^2. 
\end{align}
This way, the DDE~Eq. \eqref{normalized nonlinear forth} is reduced to the following PDE:
\begin{align}
&p_{tt}-c^2\Delta p-\frac{c^2}{\omega^2_0}
\Delta p_{tt}+\frac{1}{\omega^2_0\left(\kappa+1\right)}p_{tttt}\nonumber\\
&-\frac{b}{1+\kappa}\left(\left(p^2\right)_{tt}+\frac{1}{\omega^2_0}\left(p^2\right)_{tttt}\right)=0, \label{Boussinesq}
\end{align}
where subscripts denote partial derivatives. The above effective PDE, Eq.~\eqref{Boussinesq}, has the form of a 2D Boussinesq equation, i.e. a weakly dispersive and weakly nonlinear PDE in two dimensions, which has been used to model surface waves in hydrodynamics \cite{karpman,ablowitz}. 
  
It is useful to compare the linear dispersion relation of Eq.~(\ref{Boussinesq}) with the one derived via the TMM approach ---see Eq.~(\ref{TMM resonators}). The former, can be derived upon considering small amplitude plane wave solutions of Eq.~(\ref{Boussinesq}), namely:
\begin{equation}
  p(\mathbf{r},t)=p_0\mathrm{e}^{\mathrm{i}(\mathbf{k}\cdot\mathbf{r}-\omega t)}+{\rm c.c.}, \quad p_0\ll1 , \label{plane wave}
\end{equation}
where $\mathbf{k}=(k_x,k_y)$, 
and $\mathbf{r}=(x,y)$. This leads to  the dispersion relation:
%
\begin{equation}  
k(\omega) =\pm\frac{\omega}{c(\kappa+1)^{1/2}}\sqrt{\frac{(\kappa+1)\omega^2_0-\omega^2}{\omega^2_0-\omega^2}},
\label{boussinesq dispersion2}
\end{equation}
where $k=(k_x^2+k_y^2)^{1/2}$ and $\pm$ signs correspond to outgoing- and ingoing waves. 
%
The above dispersion relation can further be simplified in the long-wavelength approximation, i.e., for  $\omega\ll\omega_B$; indeed, upon Taylor expanding Eq.\,\eqref{boussinesq dispersion2} we obtain (for outgoing waves):
 \begin{equation}
    k=\frac{1}{c}\omega+\frac{\kappa}{2c \omega_0^2(\kappa+1)}\omega^3+\mathcal{O}(\omega^5).
    \label{lkdv}
 \end{equation}
Comparing isofrequency lines of Eqs.~(\ref{boussinesq dispersion2}) and (\ref{lkdv}) [see dotted (black) and (red) circles in Fig.~\ref{dispersion}(c)] with ones obtained via the TMM [solid (white) lines in the same figure)], we can readily observe that the continuum approximation is in excellent agreement with the TMM. 
By choosing $\omega_0\ll\omega_B$, we ensure the validity of the EA analogue and of the Boussinesq equation\,\eqref{Boussinesq}. In addition, the coupling parameter $\kappa$ should be $\kappa \lesssim 1$ to ensure that the hybridization gap and the Bragg bandgap remain decoupled, as shown in Fig.\,\ref{dispersion}(a) (for more details see also Refs. \cite{sugimoto_dispersion,sugimoto_review,vassos_soliton}).  
 
 
We now proceed by considering small-amplitude radially symmetric solutions of Eq.~(\ref{Boussinesq});  
in such a case, the Laplacian can be expressed as $\Delta= \partial_r^2+ (1/r)\partial_r$, where $r$ is the radial coordinate. As we will see, such solutions exist in the regimes of    
weak nonlinearity and weak dispersion,  and are characterized by large radii, such that the spreading due to cylindrical divergent terms is also weak. In particular, we seek solutions of Eq.~\eqref{Boussinesq} in the form of the following asymptotic expansion: 
\begin{equation}
    p=\varepsilon p_1+\varepsilon^{2} p_2+\varepsilon^{3} p_3+\cdots, \label{series1}
\end{equation}
where $0<\varepsilon \ll 1$ is a formal small parameter (setting the amplitude of the solutions), and $p_i$ are unknown functions depending on the slow variables: 
\begin{equation}
T=\varepsilon^{1/2}\left(\frac{r-r_0}{c}-t\right),\quad \mbox{ and }R=\varepsilon^{3/2}(r-r_0), \label{slow scales}
\end{equation}
where $r_0$ is the initial radius of the wave. Substituting Eq.~(\ref{series1}) into Eq.~(\ref{Boussinesq}), and using the slow variables Eq.\,\eqref{slow scales}, we obtain identities at orders $O(\varepsilon)$ and $O(\varepsilon^2)$, while at order $O(\varepsilon^3)$, we obtain the following equation:
%
%
%
%
\begin{align}
    p_{1R}+\alpha p_{1TTT}+\beta p_{1}p_{1T}+\frac{1}{2R}p_{1}=0,  \label{ckdv}
\end{align}
where 
\begin{equation}
\alpha=\frac{\kappa}{2c\omega^2_0(\kappa+1)},\quad \beta=\frac{b}{c(\kappa+1)} \label{coefficients}
\end{equation}
are the dispersion and nonlinearity  coefficients respectively. Equation\,\eqref{ckdv} is the cylindrical KdV (cKdV) equation, which is a prototypical radially symmetric model, that describes ring-shaped waves  propagating sufficiently far from the origin \cite{KO}.

\section{Linear and Nonlinear Waves}

Having derived the cKdV Eq.~(\ref{ckdv}), we will now consider the propagation of linear and nonlinear (i.e., in the form of solitons) ring-shaped waves that can be supported in the acoustic network. We will first examine linear waves of low amplitude, such that the quadratically nonlinear term in Eq.~(\ref{ckdv}) can be neglected. Then, we will study the fully nonlinear version and investigate the dynamics of ring solitons.  

\subsection{Linear ring-shaped waves -- self-similarity}

First, we aim to ascertain the behavior of linear ring-shaped waveforms in the acoustic network. For sufficiently low amplitudes, namely for a normalized pressure amplitude $p_0$ of the order of $1~{\rm Pa}$, one may omit the nonlinear term in Eq.\,\eqref{ckdv}, and study its linearized version, namely:
\begin{equation}
    p_{1R}+\alpha p_{1TTT}+\frac{1}{2R}p_1=0, 
    \label{linearized ckdv}
\end{equation}
supplemented with the boundary condition $p(R=R_0,T)=p_0(T)$. 
Motivated by the self-similar solution of the linearized KdV in 1D   \cite{ablowitz,karpman}, as well as the approximate solution of the linearized cKdV \cite{dorfman_airy},
we will seek self-similar solutions for the linearized cKdV. In our case, such solutions retain their temporal shape, and only feature spatially-dependent amplitudes and/or widths (see Ch.~5 of Ref.~\cite{logan2008}). A self-similar solution of Eq.~\eqref{linearized ckdv} may be sought in the form: 
\begin{equation}
    p_{\text{sim}}=\frac{1}{R^w}f(\eta),\quad \eta=\frac{BT}{R^q}, \label{self ansatz}
\end{equation}
where $f(\eta)$ is an unknown function of the coordinate $\eta$, 
assumed to satisfy homogeneous boundary conditions at infinity, i.e., $f(\eta)\to 0$ as $\eta\to\infty$; furthermore, 
$B$, $w$ and $q$ are parameters that will be found self-consistently, so that to warrant self-similarity. Substituting the ansatz, given in Eq.\,\eqref{self ansatz},  in Eq.~\eqref{linearized ckdv}, 
and gathering equal powers of $R$, we obtain,
\begin{eqnarray}
\frac{1}{R^{w+1}} \left[\left(w-\frac{1}{2}\right)f(\eta)+qf^\prime(\eta)\eta\right]
-\frac{\alpha B^3}{R^{w+3q}}f^{\prime\prime\prime}(\eta)=0,\nonumber \\ 
\label{powers}
\end{eqnarray}
where $f^\prime$ denotes the derivative of $f$ with respect to $\eta$. Next, to ensure self-similar behavior of the solution, first we require each term of Eq.~\eqref{powers} to be of the same power in $R$; this leads to $q=1/3$. Second, to reduce Eq.~(\ref{powers}) to an ordinary differential for $f(\eta)$ (independent of any parameter), we choose    
%
$w=5/6$ and $B=(3\alpha)^{-1/3}$. 
This way, Eq.\,(\ref{powers}) reduces to $f^{\prime\prime\prime}-(\eta f)^\prime=0$ which ---due to the homogeneous boundary conditions--- is finally reduced to the Airy equation: 
%
\begin{align}
 f^{\prime\prime}-\eta f=0.
 \label{eq airy}
\end{align}
%
%
%
The bounded solution of the Airy equation is $f(\eta)=\Ai(\eta)$, where ${\rm Ai}$ is the Airy function, which has the following integral representation:
\begin{equation}
    \Ai(\eta)=\frac{1}{2\pi}\int_{-\infty}^{+\infty}ds 
    \exp \left[\mathrm{i}\left(s\eta+\frac{s^3}{3}\right)\right].
\end{equation}
To this end, the asymptotic solution of Eq.~(\ref{linearized ckdv}) is of the form \cite{ablowitz}: $p_1(T,R) \sim p_{\rm sim}(T,R) \hat{p}_0(0)$ (where $\hat{p}_0(0)$ is the Fourier transform of the boundary condition at $R=R_0$). This suggests that the asymptotic solution of the linearized Boussinesq model, Eq.~\eqref{Boussinesq}, takes the form:
\begin{equation}
\!\!\!\!
    p(r,t)\sim \frac{\hat{p}_0(0)}{(r-r_0)^{-5/6}}\Ai(\eta), 
    \quad 
    \eta =\frac{\frac{r-r_0}{c}-t}{(3\alpha (r-r_0))^{1/3}}.
    \label{airy solution}
\end{equation}
From the asymptotic behavior of the Airy function (see, e.g., Ref.~ \cite{karpman}), one can deduce that in the limit $\eta \to 0$ the solution will vary as $p(r,t)\sim r^{-5/6}$, which is in agreement with the approximate solution found in Ref.~\cite{dorfman_airy}. Notice that, due to the curvature term ($\propto 1/r$), the decay law $r^{-5/6}$ differs from that of the linearized 1D KdV, which is $\sim r^{-1/3}$. On the other hand, in the case where the dispersion coefficient vanishes $\alpha\to 0$ the far-field amplitude decay in 2D free space is proportional to $\sim r^{-1/2}$. Then, one may notice that the joint effects of the curvature-induced decay and dispersion yield a decay rate $\sim r^{-5/6}$, which coincides with the decay rate of the self-similar solution Eq.\,\eqref{airy solution}. 

Next, we corroborate our theoretical findings, i.e the theoretical self-similarity predictions for the propagation of linear dispersive rings, with direct numerical simulations of the simplified quasi-1D conservation equations \eqref{mass simplified waveguide}-\eqref{mass simplified}. 
In the simulations, 
each waveguide segment of length $d$ is discretized to $\tilde{N}$ points (see Appendix \ref{appendix num} for details). Having in mind the possibility of future experiments, we choose to simulate a realizable  experimental platform. In particular, we excite the system with a boundary condition on a circle of radius $r_0$, located at the center of the network. The size of the network used in the simulations, is $401\times401$ unit cells. Parameter values are the ones mentioned in Section~II.
\begin{figure}[tbp!]
\begin{center}
\includegraphics[width=0.475\textwidth]{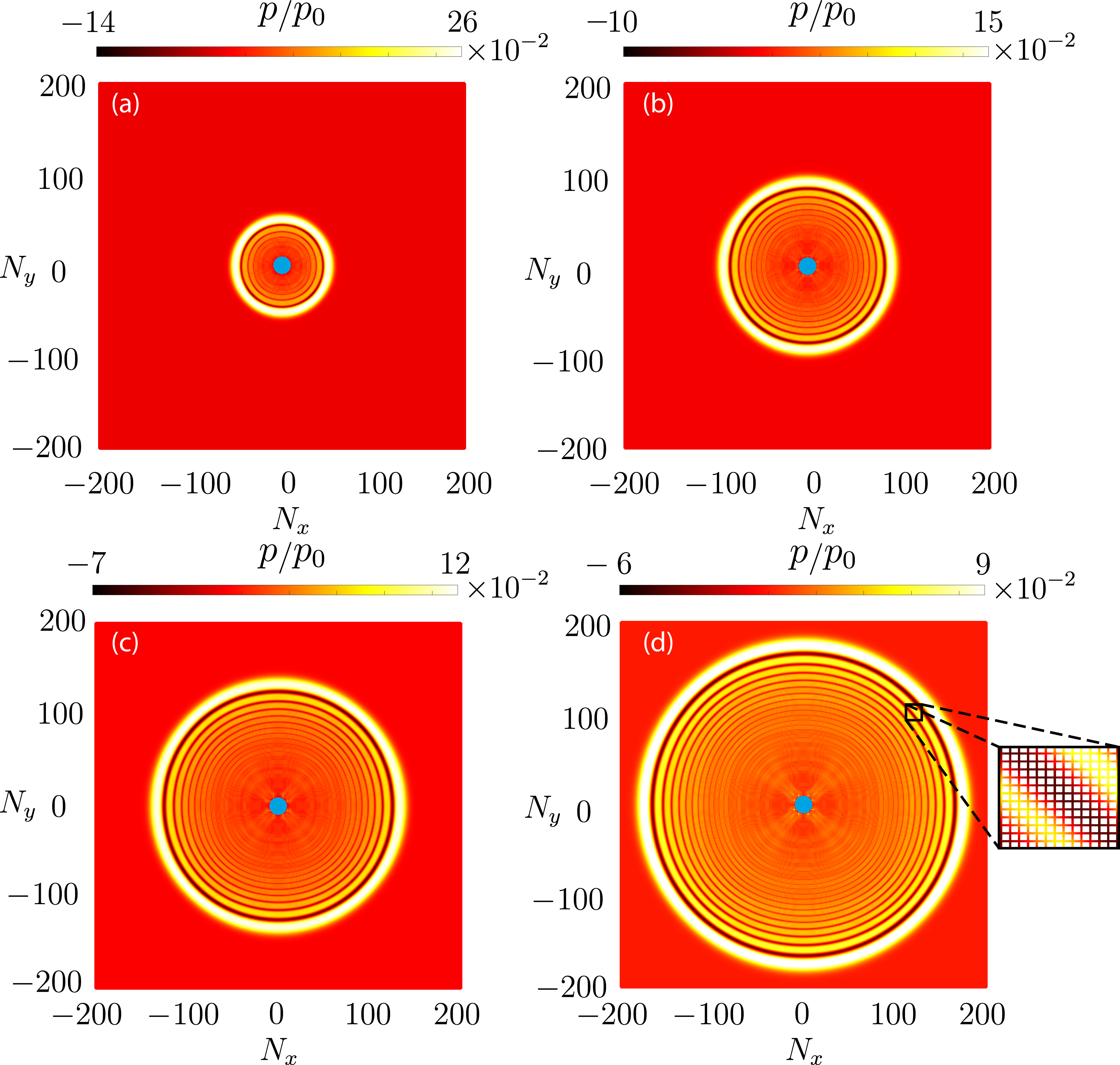}
\caption{Contour plots of the pressure field, for the low-amplitude (linear) ring-shaped self-similar waveform, at times $t=787.5,\,1575,\,2362.5,\,3150$. 
The inset in panel (d) shows a zoom of the pressure profile in the network, revealing the waveguide structure.}  \label{contour linear}
  \end{center}
\end{figure}
To compare the results of the simulations with the analytical predictions, we fix the  perturbation parameter to $\varepsilon=0.1$, while the radius of the cylindrical boundary condition is set to  $r_0=10d=1/\varepsilon$. This choice for the value of the initial radius ensures the compatibility of our numerical results, with the validity of our analytical approximation  (where sufficiently large radii where considered). Furthermore, for the boundary condition, we use a generic Gaussian pulse,  
\begin{equation}
p(r_0,t)= p_0\exp\left(-\left((t-t_0)/\sigma\right)^{2}\right),
\label{BC}
\end{equation}
of amplitude $p_0$ and width $\sigma$. We fix  $p_0=1.2 \times 10^{-5}$, corresponding to a peak pressure of $1~{\rm Pa}$ in physical units (which ensures that nonlinear effects are negligible); furthermore, we use the value $\sigma=29$, corresponding to $\approx 0.9$~ms   


Contour plots of the resulting pressure field are presented in  Fig.\,\ref{contour linear} at times  $t=787.5,\,1575,\,2362.5,\,3150$, in panels~(a)-(d) respectively. A zoom of the pressure field of the network in panel~(d) portrays the structure of connected waveguides. On the periphery of the blue disk at the center of the network, we apply the boundary condition of Eq.~(\ref{BC}), while in the interior we use a Neumann boundary condition (zero flux). 
%

\begin{figure}[tbp!]
\begin{center}
\includegraphics[width=0.475\textwidth]{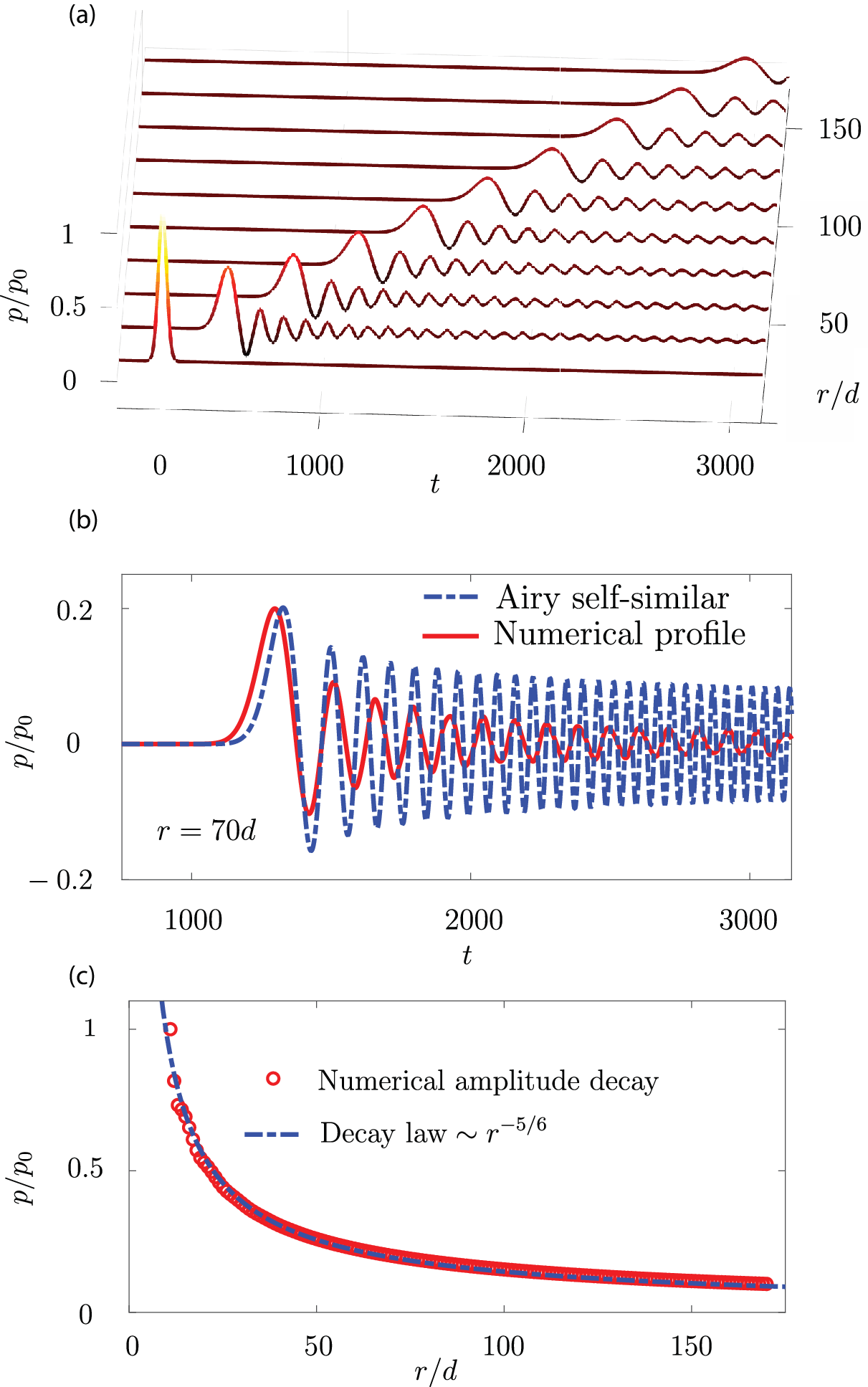}
  \caption{(a) Evolution of the initial pressure pulse along the direction $\Gamma X$ at fixed distances. (b) Comparison of the numerical simulation [solid line (red)] and the Airy self-similar solution Eq.\,\eqref{airy solution} [dotted line (blue)] at $r=70d$. (c) Amplitude decay along the direction $\Gamma X$; numerical simulation [circles (red)] and the analytical prediction Eq.\,\eqref{airy solution} [dotted line (blue)].
  \label{temporal linear}}
  \end{center}
\end{figure}

The evolution of the initial pulse along the direction $\Gamma X$ is also depicted in the 3D plot of Fig.~\ref{temporal linear}(a); notice that the evolution along the direction $\Gamma M$ (not shown here) is almost identical. 
Each of the individual snapshots, shown in this figure at fixed distances, features an increasing width due to dispersion, and 
exhibits a profile which follows the shape of the Airy function. In fact, as seen in Fig.~\ref{temporal linear}(b), a direct comparison between the numerical result [solid (red) line] and the estimation of Eq.~(\ref{airy solution}) [dashed (blue) line] at $r=70d$, shows a very good agreement between the two ---especially given the asymptotic nature of Eq.~(\ref{airy solution}). Notice that the agreement is better for the main pulse, and becomes worse for the secondary pulses, with errors  basically occurring in the amplitude and, in large times, also in the frequency of the dispersive waveform. It is finally mentioned that, as seen in Fig.~\ref{temporal linear}(a), the amplitude of the main pulse decays rapidly as it propagates in the network. The relevant decay rate (again along the direction $\Gamma X$) is shown in Fig.~\ref{temporal linear}(c). In particular, shown are the numerical result [(red) circles] and the analytically obtained decay law, $\propto r^{-5/6}$, pertinent to the self-similar solution [dashed (blue) line]. Obviously, there is an excellent agreement between the two.  
%
 %

\subsection{Ring solitons}

We now proceed with the case where the dispersion, nonlinearity, and curvature-induced decay terms of the cKdV equation~(\ref{ckdv}) are of the same order. This optimal balance may be achieved for amplitudes of the order of $\sim10$\,kPa, and initial radii of the order of $\sim10d$. 
\begin{figure}[h!]
\begin{center}
\includegraphics[width=0.475\textwidth]{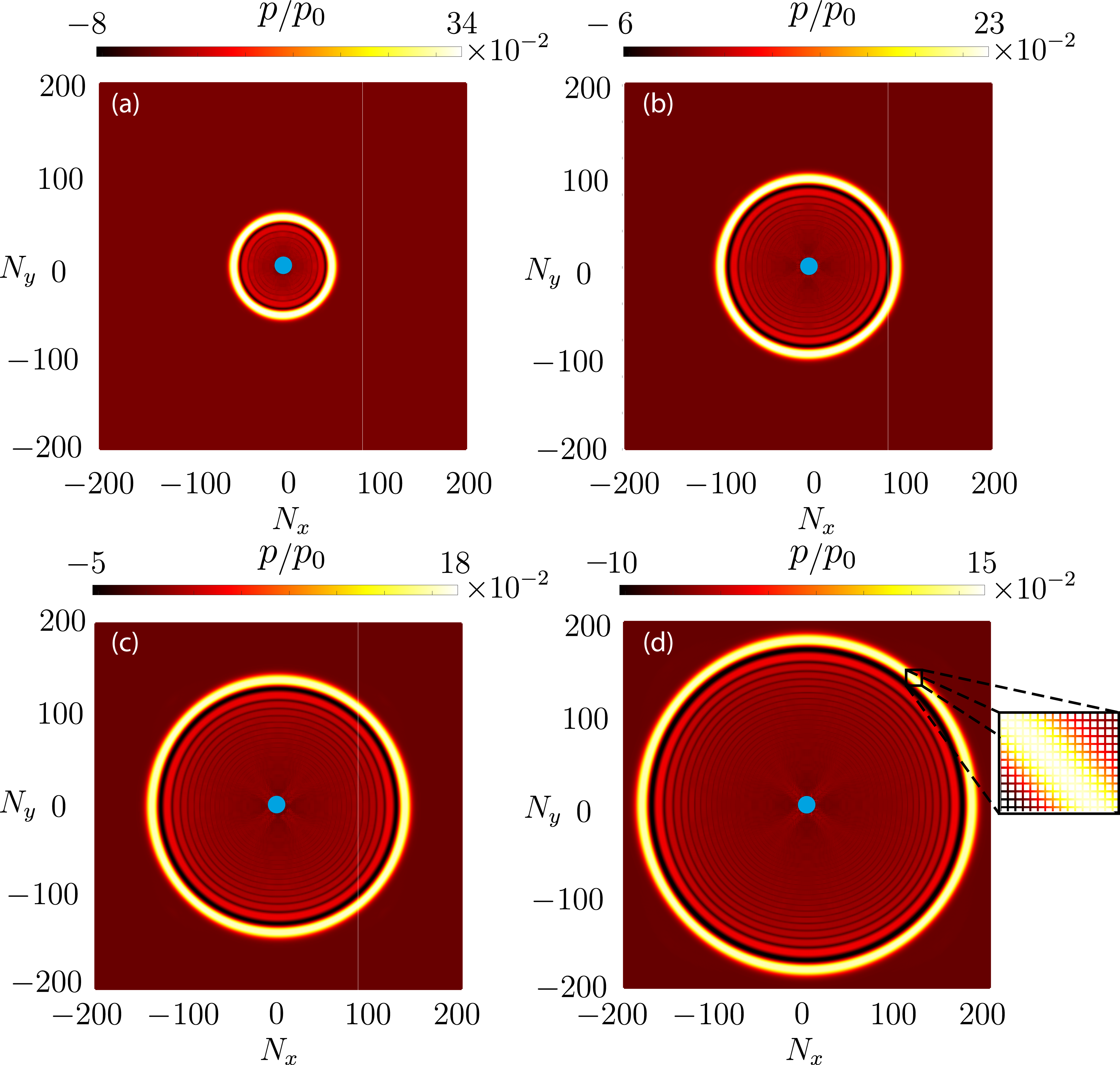}
  \caption{Contour plots of the pressure field, for the high-amplitude (nonlinear) ring soliton, at times $t=787.5,\,1575,\,2362.5,\,3150$. 
The inset in panel (d) shows a zoom of the ring soliton's pressure profile in the network, revealing the waveguide structure.  \label{contour soliton}}
  \end{center}
\end{figure}
In this case, there exist an exact ring-shaped soliton solution (usually called ``cylindrical soliton'') of the cKdV equation, which can be found by the formal reduction of the cKdV to the usual 1D KdV model \cite{hirota1979exact}. The relevant solution is a ring-shaped pulse of a $\sech^2$ profile, on top of a rational background $\propto T/R$ (see, e.g., Refs.~\cite{hirota1979exact,mannan2015ring}). Obviously, the divergence of the background at the origin (i.e., for $R\rightarrow 0$), makes the above exact solution difficult to be investigated either numerically or experimentally. Nevertheless, a much more convenient form of the cylindrical soliton is available. Indeed, an asymptotic analysis, valid for solutions of sufficiently large radii \cite{KO,johnson1999note}, shows that an approximate form of the cylindrical soliton's core can be well approximated by the planar KdV soliton, but with a slowly varying amplitude $A(R)$, due to the expansion of the solution.
\begin{figure}[tbp!]
\begin{center}
\includegraphics[width=0.475\textwidth]{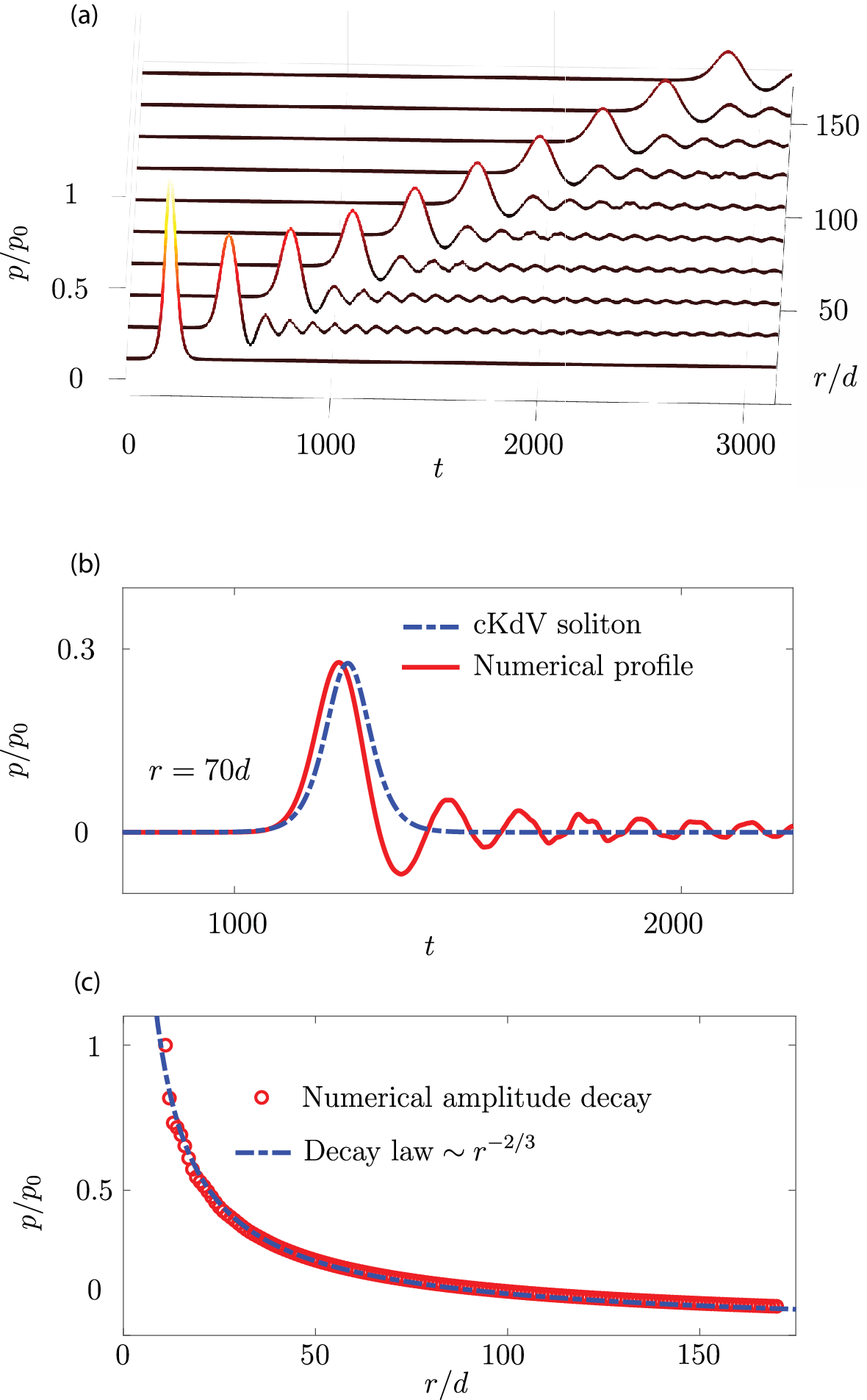}
  \caption{(a) Evolution of the initial ring soliton along the direction $\Gamma X$ at fixed distances. (b) Comparison of the numerical simulation [solid line (red)] and the cKdV soliton solution \eqref{soliton solution} [dotted line (blue)] at $r=70d$. (c) Amplitude decay along the direction $\Gamma X$; numerical simulation [circles  (red)] and the analytical prediction \eqref{soliton solution} [dotted line (blue)].\label{temporal_soliton}}
  \end{center}
\end{figure}
%
%
This approximate form of the soliton is: 
\begin{align}
&p_1(T,R)\approx A(R)\sech^2{\left[w_0(R_0)\left(T-\frac{R-R_0}{v(R_0)} \right)\right]},\label{soliton approximate}
\end{align}
where $A(R)=A_0\left(R_0/R\right)^{2/3}$ is the spatially-varying soliton amplitude (with $A_0$ being the soliton amplitude at the initial radius $R=R_0$), while the soliton's width $w_0(R)$ and velocity $v(R)$ are accordingly given by:
\begin{eqnarray}
w_0(R)=\left(\frac{A(R)\beta}{12\alpha}\right)^{1/2}, 
\quad 
v(r)=\frac{3}{A(R)\beta}.
\end{eqnarray} 
It is convenient to express Eq.~(\ref{soliton approximate}) in terms of the original variables $r$ and $t$, as follows:
\begin{eqnarray}
p(r,t) &\approx& \varepsilon A(r)  
\sech^2 \Big\{ \varepsilon^{1/2}w_0(r)
\nonumber \\ 
&\times&
\left[\left(\frac{1}{c}-\varepsilon\frac{1}{v(r)}\right)\left(r-r_0\right)-t\right]\Big\}. \label{soliton solution}
\end{eqnarray}
%
%

To verify the existence of the cylindrical soliton in the acoustic network, we proceed by presenting results of numerical simulations of Eqs.\,\eqref{mass simplified waveguide}-\eqref{mass simplified}, as we did for the linear case. In this case, the boundary condition at $r=r_0=10d$ is chosen to be a $\sech^2$ pulse, whose amplitude-width relation is given by the cKdV approximate solution given by Eq.\,\eqref{soliton solution}, namely:
\begin{equation}
    p(r_0,t)=\varepsilon A_0\sech^2{\left(\varepsilon^{1/2}w(r_0)t\right)}, \label{soliton condition}
\end{equation}
where the normalized amplitude $A_0$ is chosen so that it corresponds to a peak pressure of 27.25\,kPa; this, subsequently fixes the initial inverse width $w(r_0)$, and the initial velocity $v(r_0)$ of the pulse, respectively.

Similarly to the linear case, Fig.\,\ref{contour soliton} depicts contour plots of the resulting pressure field at times  $t=787.5,\,1575,\,2362.5,\,3150$ ---see panels~(a)-(d) respectively. A zoom of the pressure field in panel~(d) illustrates the localized nonlinear wavefront of the ring soliton in the network. Once again, the periphery of the blue disc depicts the location of the boundary condition \eqref{soliton condition} [Neumann boundary condition (zero flux) are taken in the interior of the disk].


The evolution of the initial pulse along the direction $\Gamma X$ is also depicted in the 3D plot of Fig.~\ref{temporal_soliton}(a); notice that similarly to the linear case, the evolution along the direction $\Gamma M$ (not shown here) is almost identical. 
Each of the individual snapshots, shown in this figure at fixed distances, features an amplitude decay of the soliton's core, while its width slowly increases due the balance of dispersion and nonlinearity, as predicted.

Notably, in panel\,(b) we corroborate our numerical result [solid (red) line] with the temporal profile suggested by the analytical prediction, Eq.\,\eqref{soliton solution} [dotted (blue) line] for $r=70d$; the two are found to be in very good agreement, especially as concerns the soliton's core. 
In particular, the numerical solution [solid (red) line in Fig.~\ref{temporal_soliton}(b)] has a structure which is composed by the main pulse ---which is well approximated by the soliton profile--- and a radiation tail. The formation of the latter, is attributed to the fact that the sech$^2$-pulse is not an exact solution of Eqs.~(\ref{mass simplified waveguide})-(\ref{mass simplified}).
Finally, the small discrepancy between the velocity of the analytical prediction and the numerical solution is of order $O(\varepsilon^2)$, which is consistent with our perturbation method.
%

The amplitude decay (for the direction $\Gamma X$) of the soliton core is presented in Fig.~\ref{temporal_soliton}(c). Here, we compare the numerical result [(red) circles] and the analytically obtained decay law of the ring soliton, $\propto r^{-2/3}$, [dashed (blue) line]. Once again, the agreement between the numerical and analytical result is excellent, solidifying the formation and propagation of ring solitons in the acoustic network.
   

\section{Conclusions}

In this work, we have studied the formation and dynamics of ring-shaped linear and nonlinear waves (solitons) that can be supported by a square lattice of acoustic waveguides. First we studied the linear dispersion relation of the acoustic network, and found that it features an anisotropic behavior. We demonstrated that the relevant anisotropy can be suppressed, in the long-wavelength limit, upon introducing Helmholtz resonators at the junctions of the network. This suppression was shown to be necessary for the formation of radially symmetric, linear and nonlinear, waveforms in the network.  


Utilizing the fluid conservation laws, we  introduced the electroacoustic analogy for the 2D network, which we then used to derive an effective 2D Boussinesq equation. The latter,  was then reduced to a cylindrical KdV (cKdV) equation, which captures the weakly nonlinear and weakly dispersive characteristics of the network. The effective cKdV description, which is valid for radially-symmetric waves of large radii, allowed us to investigate both the low- and high-pressure regimes, with the former (latter) being relevant to linear (nonlinear) waves in the network. 

We have shown that, in the linear regime (relevant to low initial pressure), the linearized cKdV model admits a self-similar solution of the form of an Airy function. This way, we have found that initial data of the form of ring-shaped Gaussian pulses undergo a dispersion-induced broadening, together with a curvature-induced decay. In fact, the decay was found to be  significantly faster than that of linear waves in free space, due to the combined effects of dispersion and curvature. Direct numerical simulations, performed at the level of the fluid equations, corroborated our analytical predictions. In particular, we have found that the profile of the main pulse, as well as its amplitude decay and its velocity, can be approximated quite accurately by the solution of the linearized cKdV.  


On the other hand, in the nonlinear regime (relevant to high initial pressure), we have predicted the existence of ring ({\it alias}  cylindrical) solitons in the acoustic network. These structures are, in fact, solutions of the cKdV equation, which decay significantly slower than the linear Airy solutions, due to the balancing of the combined effects of dispersion and curvature with nonlinearity. The theoretical predictions, obtained in the framework of the cKdV equation, were found be in very good  agreement with results of direct numerical simulations for the acoustic network.

%
%

Our analysis and findings suggest a number of interesting future research directions. In particular, while in this work we  considered a square acoustic lattice, it would be quite relevant to investigate other lattice symmetries. This way, one could 
predict novel linear and nonlinear propagation phenomena, due to  other, geometry-induced, dispersive behaviors. 
On the other hand, potential use of resonant elements other than Helmholtz resonators, and possibly along different directions of propagation, could significantly alter the dispersion of the setup. Such settings may allow for the formation of 2D localized pulses (``sound bullets'') that can be excited in the network.
%
%
It would also be particularly interesting to study wave dynamics in the fully anisotropic square network, so as to investigate (and possibly manipulate) the form of linear or nonlinear wave structures that can be formed along different directions.   
Finally, pertinent studies, also combined by experimental realizations, are of great interest.

\appendix
\section{Transfer matrix method}\label{appendixtmm}
Let us consider the unit cell, of the infinite square network, which is depicted in Fig.\,\ref{resonators system}\,\textbf{(c)}.
Our starting point is the consideration of an ideal fluid neglecting nonlinearity, viscosity and other dissipative
terms. In this regime, the acoustic pressure field $p(x,y)$ inside each waveguide of Fig. \ref{resonators system}\,\textbf{(a)} is governed by the two-dimensional (2D) Helmholtz equation:
\begin{equation}
\frac{\partial^{2} p}{\partial x^{2}}+\frac{\partial^{2} p}{\partial y^{2}}+k^{2} p=0,
\label{helmholtz}
\end{equation}
 with Neumann conditions corresponding to zero normal velocity at the rigid walls $\partial_n p=0$. 
Here $k= \omega /c_0$ with $\omega$ the
angular frequency and $c_0$ the speed of sound.
Moreover, we are interested in sufficiently low frequencies, i.e the wavelength is much larger than the cross section of each waveguide, such that only the fundamental acoustic mode of the waveguide is propagating (monomodal approximation).

We start by considering the conservation of flux in the central junction of the unit cell
\begin{equation}
u_{n, m}^{n-1, m}+u_{n, m}^{n, m+1}+u_{n, m}^{n, m-1}+u_{n, m}^{n+1, m}+u^{H}_{n,m}=0, \label{flux conservation resonators}
\end{equation}
where $u_{n, m}^{n\pm1, m+\pm1}$, $u^{H}_{n,m}$ are the incoming fluxes from the nodes $(n\pm1, m+\pm1)$ to the node $(n,m)$ and the flux entering the resonator respectively. For sufficient low frequencies one may approximate the HR as a point scatterer with the following impedance,
\begin{equation}
Z_{HR}\approx\mathrm{i}\frac{\rho_0 l}{S_n}\frac{\omega^2-\omega_0^2}{\omega}, \label{impedance resonator}
\end{equation}
with $\omega_0=c_0(S_n/(l hS_c))^{1/2}$ the resonance frequency of the HR in the longwavelength regime \cite{kinsler,hirschberg,vassos_soliton}. Notice, that the resonance frequency depends on the geometrical characteristics of the resonator, namely $l$, $h$, $S_n$, $S_c$ the radius of the neck, the height of the cavity and the cross sections of the neck and the cavity respectively.
In addition, we can express each flux as a function of the  pressure of the unit cell, $p_{n,m}$ , and the pressure of the neighbour unit cells, $p_{n\pm1, m+\pm1}$, through the TMM \cite{depollier,2dacoustic,acousticgraphene}.
After, some algebra we arrive at the following eigenvalue problem for the pressure field $p_{n,m}$,
\begin{align}
   4\left(\cos{(kd)}+\mathrm{i}\frac{Z_w}{Z_{HR}}\sin{(kd)} \right)p_{n,m}=p_{n-1,m}\nonumber\\
   +p_{n+1,m} +p_{n,m-1} +p_{n,m+1} \label{eigenvalue resonators}
\end{align}
Finally, to obtain the corresponding dispersion relation, we seek solutions in form of Bloch waves, 
\begin{equation}
p_{n, m}=p_0 e^{i \mathbf{q} \cdot\mathbf{R}_{n, m}}=p_0 e^{i q_{x} n d} e^{i q_{y} m d},
\end{equation}
where $q_x$ and $q_x$ are the Bloch wavenumber along the directions $x$ and $y$ of the first Brillouin zone.
By substituting the periodic wave solution in Eq. \eqref{eigenvalue resonators} we obtain Eq.\,\eqref{TMM resonators}. In the absence of resonators, the dispersion relation Eq.\,\eqref{TMM resonators}, is reduced to the form given in Eq.~\eqref{TMM no resonators}.

\section{Numerical scheme}\label{appendix num}
Our numerical scheme is based on the integration of the simplified fluid conservation laws, Eqs.\,\eqref{mass simplified waveguide}-\eqref{mass simplified}, with a finite-difference method. It is worth to note that the discretization of the aforementioned equations is equivalent to the electroacoustic analogy proposed in Sec.\,II (see also Refs\,\cite{lissek,vassos_soliton}). As explained in the main text, the choice of the discretization is the lattice distance $d$, which finally lead to the DDE~\eqref{nonlinear forth order}. 

For our numerical scheme, since we are interested in accurately capturing the dispersive and nonlinear characteristics of the network, we choose a \textit{smaller} discretization distance, $\tilde{d}=d/\tilde{N}$  where $\tilde{N}$ is the number of points per unit cell, for a single waveguide segment (see details in Ref.~\cite{sougleridis2023acoustic}). In terms of the electroacoustic analogue, this approach employs a transmission line, whose unit cell is in fact a ``supercell''; notice that this is a generic scheme that is commonly used in numerical studies of continuum systems \cite{deymier}. 
For the numerical simulations presented in this study we fix $\tilde{N}=10$.

\bibliography{biblio.bib}

\end{document}